\documentclass[lettersize,journal]{IEEEtran}
\usepackage{amsmath,amsfonts}
\usepackage{algorithmic}
\usepackage{algorithm}
\usepackage{array}
\usepackage[caption=false,font=normalsize,labelfont=sf,textfont=sf]{subfig}
\usepackage{textcomp}
\usepackage{stfloats}
\usepackage{url}
\usepackage{verbatim}
\usepackage{graphicx}
\usepackage{cite}
\usepackage{makecell}
\begin{document}

\title{Test Primitive:A Straightforward Method To Decouple March}

\author{~\IEEEmembership{Yindong Xiao, Shanshan Lu, Ensheng Wang, Ruiqi Zhu, Zhijian Dai}
\thanks{}}

\IEEEpubid{
}

\maketitle

\begin{abstract}
The academic community has made outstanding achievements in researching the March algorithm. However, the current fault modeling method, which centers on fault primitives, cannot be directly applied to analyzing the March algorithm. 
This paper proposes a new test primitive.The test primitives, which decouple the cell states from sensitization and detection operations, describe the common features that must be possessed for the March algorithm to detect corresponding faults, forming a highly flexible and scalable March algorithm analysis unit. The theoretical analysis proves that the test primitives demonstrate completeness, uniqueness, and conciseness. On this foundation, the utilization of test primitives within the March analysis procedure is elucidated.
\end{abstract}

\begin{IEEEkeywords}
Memory Test,March Algorithm,Fault Primitive,Test Primitive.
\end{IEEEkeywords}

\section{Introduction}
\IEEEPARstart{M}{arch}  algorithms, as a class of test vector generation algorithms, are widely used in the field of memory testing. Due to their excellent test efficiency and high fault coverage , they have been extensively studied by scholars at home and abroad \textsuperscript{\cite{1}}. Understanding of March algorithms in the industry is continuously deepening and becoming more abstract. Each new perspective and high-level theory has brought progress at the application level.In recent years, Jidin et al. proposed the cell trend theory based on address direction and the state of the previous memory cell \textsuperscript{\cite{2}}. They applied this theory to non-connectivity static fault coverage analysis and automatic generation of March algorithms, achieving significant progress \textsuperscript{\cite{3,4}}. However, dealing with dynamic faults brought by technologies such as FinFET remains challenging.

On the other hand, changes in application-level demands have also guided the direction of theoretical research on March algorithms. With the increasing focus of industry on hierarchical memory diagnosis techniques \textsuperscript{\cite{5}}, the need for scalable and flexible test primitives has arisen. In 2000, Van De Goor et al. \textsuperscript{\cite{6}}proposed fault primitives to represent various behavioral-level faults. This model has been widely used to describe memory faults. Building on this, Al-Ars et al. proposed the concept of test primitives with the goal of developing a scalable and flexible model dedicated to fault detection. This was proposed to address the issue that fault primitive models cannot be directly used for test diagnosis.However, the test primitive theory proposed by Al-Ars et al. \textsuperscript{\cite{7}} primarily focuses on ensuring uniqueness and conciseness of primitives. The feature of coupling memory cells  states with sensitization and detection operations makes proving completeness difficult. Therefore, this theory cannot be applied to algorithm analysis and generation tasks.

Based on fault primitives, this paper defines new test primitives to describe the characteristics of memory access operations required to detect corresponding faults. The proposed test primitives demonstrate completeness, uniqueness, and conciseness. This paper examines the construction method for a test primitive model and establishes a test primitive library that covers non-connectivity static faults. The functional-level fault model of memory can be linked with the test primitives, simplifying analysis of the fault coverage of March algorithms.

Completeness: The test primitives encompass all characteristics of March algorithms that can detect corresponding faults.

Uniqueness: The test primitives do not include descriptions that cannot test corresponding faults, ensuring one fault primitive corresponds to only one test primitive.

Conciseness: When the test primitives cover corresponding faults, the number of detection operations is minimal without introducing redundant operations.

The content of this paper is organized as follows: Section 2 mainly introduces the target fault types and their test primitives. Section 3 proves the properties of the proposed test primitives. Section 4 elucidates the utilization of test primitives within the March analytical procedure.Finally, the full text is summarized.

\section{Fault Model and Its Test Primitives}
\subsection{Fault Model}
The fault model of memory can be divided into static and dynamic faults based on the number of operations required to sensitize the fault. Static faults fall into simple faults and connectivity faults depending on whether faults affect one another. While primarily considering non-connectivity static faults encompassing single-cell faults as well as coupled faults, which implicate multiple memory cells, this paper exclusively examines the scenario of two memory cells, namely double-cell coupled faults that represent a type of memory fault residing autonomously in a single memory cell without influencing other cells, as opposed to double-cell faults designating two coupled memory cells commonly termed aggressor cells (a) and victim cells (v). Common single-cell faults and double-cell coupled faults appear in Table \ref{tab:table2}.The fault primitive expression of a single-cell fault is $\left\langle S/F/R \right\rangle$, where $S$ represents the operation or state that triggers the memory fault behavior, $F$ represents the state value of the faulty cell, and $R$ represents the logic output level of the read operation.The fault primitive expression of a double-cell fault is $\left\langle Sa; Sv/F/R \right\rangle$, Where $Sa$ and $Sv$ represent the state or operation of the aggressor cell and the victim cell that trigger the memory fault, respectively.$Sa$ and $Sv$ cannot refer to operations simultaneously \textsuperscript{\cite{8}}.

\subsection{Test Primitive  Definition}
With the continuous improvement of memory manufacturing technology, many new types of faults have emerged, requiring the March algorithm to continuously adapt to new types of faults. The high test complexity of the test algorithm makes the analysis, verification, and generation of the algorithm difficult. Based on fault primitives, this paper decouples the cell state in the March algorithm from fault sensitization and detection, establishes a test primitive , and simplifies the analysis, verification, and generation process of the March algorithm.

The test primitive  is defined to describe the common characteristics of the March algorithm that can detect the corresponding fault primitive. It can be obtained directly from the fault primitives, and each fault primitive corresponds to a test primitive. The definition of the test primitive is based on the fault primitive, described in the form of$\left\langle S \right\rangle D$, which describes the characteristics of the operations required to detect the corresponding fault, where $\left\langle S \right\rangle$ is called the sensitization operation set, and $D$ is the detection operation feature descriptor, used to describe the detection operation set features that can verify the cell state after sensitization. Similarly, the double cell fault $\left\langle Sa;Sv/F/R \right\rangle$ can be detected using the test primitive in the form of $\left\langle Sa;Sv \right\rangle D$.

Test primitives and fault primitives have the same sensitization operation set. Therefore, the key to deriving test primitives is studying the applicability of detection operation feature descriptors. The detection operation feature descriptor includes the read operation for detecting faults, hereinafter referred to as the detection operation, and special symbols describing the position of the detection operation. Symbols similar to regular expressions are used to indicate the position of the detection operation. Common symbols and their meanings are shown in Table \ref{tab:table1}.The detection descriptor is divided into two cases according to the address change direction. ``a$\Rightarrow$v" indicates that the address change direction is from the aggressor cell to the victim cell. ``v$\Rightarrow$a" indicates the opposite. Before proposing this representation method in this paper, the forms ``a$>$v" and ``a$<$v" were used to indicate the relative address relationship between the aggressor cell and the victim cell.When generating the test primitive of the fault primitive, the relative address relationship between the aggressor cell and the victim cell needs to be considered in both the increasing address direction and the decreasing address direction. Using the ``a$\Rightarrow$v" and ``v$\Rightarrow$a" methods proposed in this paper combines the address change direction and the relative address relationship between the victim cell and the aggressor cell, making the analysis process more convenient.

\begin{table}[!t]
\caption{ Common Symbols and Their Meanings in Test Primitives \label{tab:table1}}
\centering
\begin{tabular}{|c|c|}
\hline
Symbol & Meaning\\
\hline
$*$	& \makecell{Indicates that 0 or more arbitrary memory access \\operations can be added.}\\
\hline
 \#	& \makecell[c]{Indicates that the sensitization and detection of faults can\\ be split into two March elements at the current position,\\ but this split is not required.If a split is not chosen, \\the address traversal order before and after "\#" \\remains consistent.}\\
\hline
;& \makecell{Indicates that fault detection and fault sensitization \\ be split into two March elements at the current position.}\\
\hline
\^{} & \makecell{Indicates that the memory access operation following it\\ must be in the first position of the March element,and \\fault sensitization is after the detection operation}\\
\hline
\end{tabular}
\end{table}

According to the following method to derive the test primitive, where the fault cell of the double cell coupled fault refers to the victim cell:

\begin{itemize}
  \item [1.] 
  If the fault cell executes a read operation during sensitization, and the expected state of this read operation differs from R in the fault primitive, $D =\phi$. Otherwise, there must be a detection operation in the detection operation descriptor.  
  \item [2.]
  The expected state of the detection operation is determined by the state achieved by sensitizing the fault cell, i.e. the state after the sensitization operation or the sensitization state itself.
  \item [3.]
 For a single cell fault,``\#" is added before the detection operation, regardless of whether the sensitization of the fault cell involves a state or an operation.
 \item[4.] 
 Double cell faults are divided into the following three cases according to the different methods of sensitizing the aggressor cell and the victim cell:
 \begin{itemize}
     \item [a.]If the aggressor cell executes an operation during sensitization, add ``\^{}" before the detection operation in the a$\Rightarrow$v case. Add ``*" after the detection operation. Add ``;" before the detection operation in the v$\Rightarrow$a case. Add ``*" after the sensitization operation.
     \item [b.]If the victim cell executes an operation during sensitization, the detection operations come after the sensitization operation in both the a$\Rightarrow$v and v$\Rightarrow$a cases. ``\#" is added before the detection operation.
     \item [c.]If both the aggressor cell and the victim cell are in the state during sensitization, two cases must be distinguished. The aggressor cell and the victim cell can separately occupy the current address cell (CAC). When the aggressor cell occupies the CAC, insert ``\^{}"  before the detection operation and ``*"  after it. When the victim cell occupies the CAC, the detection operation follows sensitization.
 \end{itemize}
 
\end{itemize}

Here, it needs explanation that if the aggressor cell executes an operation during sensitization, the different positions of the detection operation in the a$\Rightarrow$v and v$\Rightarrow$a directions ensure that the detection operation lags behind fault sensitization in time. The above test primitive derive method also applies to dynamic faults because detection of dynamic faults is also completed by read operations after ensuring the fault sensitization conditions are met. As long as its detection operation meets the conditions for generating the above rules, its test primitive can also be easily express.

\begin{table}[!t]
\caption{Test Primitive Library\label{tab:table2}}
\centering
\begin{tabular}{|c|c|c|}
\hline
Fault Model &	Fault Primitive &	Test Primitive\\
\hline
SF& $\left\langle x/\overline{x}/-\right\rangle$ &$\left\langle x\right\rangle \#Rx$\\
\hline
SAF &	$\left\langle \forall/x/-\right\rangle$ &	$\left\langle x\right\rangle \#Rx$\\
\hline
TF & $\left\langle xW\overline{x}/x/- \right\rangle$ & $\left\langle xW\overline{x}\right\rangle \#R\overline{x}$ \\
\hline
WDF & $\left\langle xWx/\overline{x}/-\right\rangle$ & $\left\langle xWx\right\rangle \#Rx$ \\
\hline
RDF &$\left\langle Rx/\overline{x}/\overline{x}\right\rangle$ & $\left\langle Rx\right\rangle$ \\
\hline
DRDF & $\left\langle Rx/\overline{x}/x\right\rangle$ & $\left\langle Rx\right\rangle \#Rx$ \\
\hline
IRF & $\left\langle Rx/x/\overline{x} \right\rangle$ & $\left\langle Rx\right\rangle$ \\
\hline
CFst & $\left\langle x;y/\overline{y}/-\right\rangle$ & $\left\langle x;y\right\rangle \left\{
\begin{array}{cc}
    Ry &  (v\enspace in\enspace CAC)\\
     \{^{}Ry& (a\enspace in \enspace CAC)
\end{array}
\right.
$ \\
\hline
CFdsrx & $\left\langle Rx;y/\overline{y}/-\right\rangle$ & 
$\left\langle RX;y\right\rangle \left\{
\begin{array}{cc}
    \{^{} Ry* & (a\Rightarrow v) \\
    *;Ry & (v\Rightarrow a) 
\end{array}
\right.
$ \\
\hline
CFdsxwx & $\left\langle xWx;y/\overline{y}/-\right\rangle$ &
$\left\langle xWx;y\right\rangle \left\{
\begin{array}{cc}
    \{^{}Ry* &  (a\Rightarrow v)\\
    *;Ry & (v\Rightarrow a)
\end{array}
\right.
$\\
\hline
CFdsxw!x & $\left\langle xW\overline{x};y/\overline{y}/-\right\rangle$ & $\left\langle xW\overline{x};y\right\rangle \left\{
\begin{array}{cc}
    \{^{}Ry* &(a\Rightarrow v)  \\
     *;Ry &(v\Rightarrow a) 
\end{array}
\right.
$\\
\hline
CFtr & $\left\langle x;yW\overline{y}/y/-\right\rangle$ & $\left\langle x;yW\overline{y}\right\rangle\#R\overline{y}$ \\
\hline
CFwd & $\left\langle x;yWy/\overline{y}/-\right\rangle$ & $\left\langle x;yWy \right \rangle\#Ry$\\
\hline
CFrd & $\left\langle x;Ry/\overline{y}/\overline{y}\right\rangle$ & $\left\langle x;Ry\right\rangle$ \\
\hline
CFdrd & $\left\langle x;Ry/\overline{y}/y\right\rangle$ & $\left\langle x;Ry\right\rangle \#Ry$ \\
\hline
CFir & $\left\langle x;Ry/y/\overline{y}\right\rangle$ &$\left\langle x;Ry\right\rangle$ \\
\hline

\multicolumn{3}{|c|}{
S
\makecell[l]{
Please note:\\ 1) \enspace For the CFdsrx fault, when $x=y$, detection operation is empty. \\However, the sensitized read operation must be in the first position \\ of the March element.  According to the above description, its test \\ primitive can be expressed as $\left\langle Rx; y \right\rangle(x=y)$ .\\2)\enspace  For the CFst fault, when the victim cell occupies the CAC,\\ another test primitive for $x=y$ is $\left\langle x;y \right\rangle ;Ry$.
}
}\\
\hline
\end{tabular}
\end{table}

For example, the fault primitive of Incorrect Read Fault is $\left\langle Rx/x/\overline{x} \right\rangle$, which shows that the sensitization of the fault cell is a read operation. The state after the read operation differs from R in the fault primitive, so $ D = \phi$. Its sensitization operation set is $\left\langle Rx \right\rangle$. Therefore, the test primitive of the fault primitive $\left\langle Rx/x/\overline{x} \right\rangle$ is $\left\langle Rx \right\rangle$.

For cases where the aggressor cell executes an operation during sensitization, generating the test primitive is more complex. The fault primitive$\left\langle Rx; y/\overline{y}/- \right\rangle$  is used as an example of Disturb Coupling Fault to illustrate. The sensitization condition set of  $\left\langle Rx; y/\overline{y}/- \right\rangle$ is $\left\langle Rx; y \right\rangle$. The sensitization of the aggressor cell is the operation $Rx$. The sensitization of the victim cell is the state $y$. Therefore, the detection operation is $Ry$. Because the aggressor cell executes an operation during sensitization, after the sensitization operation, 0 or more arbitrary memory access operations can be added without affecting fault detection. Add ``*". It should be noted here that the final state after adding multiple memory access operations should remain the same as the state before sensitizing the victim cell. According to the above rules, the expression of D can be obtained.

\begin{equation}
\label{label9}
D = \left\{
\begin{array}{cc}
    \{^{}Ry* & (a\Rightarrow v) \\
    *;Ry &(v\Rightarrow a) 
\end{array}
\right.
\end{equation}

Its test primitives are shown in Table \ref{tab:table2}. According to the above rules, the test primitive library of the fault set involved in this paper is recorded in Table \ref{tab:table2}.

\section{Test Primitive Nature Proof}
The test primitives proposed in this paper demonstrate completeness, uniqueness, and conciseness for the target fault primitives.If the test primitives lacked completeness, they would not guarantee the ability to analyze fault coverage for any March algorithm. Uniqueness ensures that when a certain March algorithm satisfies the test primitive description, it must detect the corresponding fault. Conciseness ensures that the test primitive can directly optimize test sequence generation.

The test primitive adds at most one read operation based on the sensitization operation, so conciseness is easier to guarantee. To address the completeness problem of the test primitive, this paper adopts proof by contradiction to prove the assumption that a certain test primitive cannot cover the March algorithm, but the algorithm can detect the fault corresponding to the original language. The existence of this kind of algorithm is analyzed. If it does not exist, the completeness of the test primitive can be proved.The uniqueness of the test primitive only needs to ensure the following three conditions simultaneously to prove its uniqueness:
\begin{itemize}
    \item [1)]The sensitization operation set of test primitive is the same as the sensitization operation set of the fault primitive.
    \item [2)] The detection operation lags behind the sensitization operation in time.
    \item[3)] After the sensitization operation completes and before the detection operation, no operation can change the state of the fault cell.
\end{itemize}
When deriving the test primitive, its uniqueness is proved; thus, it will not be repeated here.

The March algorithm comprises multiple March elements M. Each March element contains one of \{$\Uparrow$,$\Downarrow$,$\Updownarrow$\}representing address change direction and the memory access operation set$O=\{op0, op1, ...\}$, $opn \in \{W0,W1,R0,R1\}$.

As defined by the detection operation
feature descriptor, when $D \neq\phi$, the relationship between operation set $Od$ in $D$ and sensitization operation set $Os$ may include:

\begin{equation}
\label{label1}
;[O1]Od[O2]Os[O3] 
\end{equation}

\begin{equation}
\label{label2}
Od[O1];[O2]Os
\end{equation}

\begin{equation}
\label{label3}
Os[O1]Od[O2]
\end{equation}

\begin{equation}
\label{label4}
Os[O1];[O2]Od
\end{equation}

In the formula,``;" signifies the end of one march in the March algorithm, and ``[...]" indicates optional content within brackets. The above four formulas describe 20 possible positional relationships. In addition to $D=\phi$, only 21 cases require discussion.

Among the above 21 relationships between detection and sensitization operations, assume that $O1$ exists in formula \ref{label1}—that is, $;O1Od[O2]Os[O3]$—if $O1$ is a read operation and the expected state from the read operation differs from the fault cell state, such a read-write sequence does not comply with March rules. If $O1$ is a read operation and its expected state matches the fault cell state, $O1$ is $Od$. If $O1$ contains write operations, the write operation will change the fault cell state, preventing fault detection. Therefore, formulas containing $O1$ do not satisfy test primitive conditions. Similarly for \ref{label4} when $O2$ exists. No formula in \ref{label2} satisfies the condition of detection operations lagging behind sensitization operations in time; thus, they do not meet test primitive conditions. In summary, we simplify the positional relationship to:

\begin{equation}
\label{label5}
;Od[O2]Os[O3] 
\end{equation}

\begin{equation}
\label{label6}
Os[O1]Od[O2]
\end{equation}

\begin{equation}
\label{label7}
Os[O1];Od
\end{equation}

Through the above formula, adding $D=\phi$ leaves only 11 remaining combinations to consider. Next, the completeness of the test primitives for the faults addressed in this paper is demonstrated through proof  by contradiction.

\subsection{Proof of Completeness for Single Cell Fault Testing Primitives}
The fault primitive for State Fault is $\left\langle x/\overline{x}/- \right\rangle$. The sensitization operation set is $\left\langle x \right\rangle$, and the detection operation is $Rx$. According to Table \ref{tab:table2}, its test primitive is $\left\langle x \right\rangle \# Rx$ . Assume that in addition to $\left\langle x \right\rangle \# Rx$ , other test primitives can detect  $\left\langle x/\overline{x}/- \right\rangle$. The positional relationship between  sensitization operation and detection  operation contains:
\begin{itemize}
    \item [1.]Formula \ref{label5} contains four cases: (1) $O2$ and $O3$ exist; (2) $O2$ exists, $O3$ does not; (3) $O2$ does not exist, $O3$ exists; (4) Neither $O2$ nor $O3$ exist. The positional relationships for these cases are: (1) $;RxO2\left\langle x \right\rangle O3$, (2)$;RxO2\left\langle x \right\rangle$,(3)$;Rx\left\langle x \right\rangle O3$,(4)$;Rx\left\langle x \right\rangle O3$. These cannot guarantee detection operation after sensitization operation, so they cannot detect State Fault.
    \item[2.] Formula \ref{label6} contains four positional relationships: (1) $O1$ and $O2$ exist, $\left\langle x \right\rangle O1RxO2$, (2) $O1$ exists, $\left\langle x \right\rangle O1Rx$, (3) $O2$ exists, $\left\langle x \right\rangle RxO2$, (4) Neither exists, $\left\langle x \right\rangle Rx$. When $O1$ exists, multiple memory accesses operation after sensitization may change the fault cell state, so (1) and (2) cannot detect State Fault. However, (3) and (4) can. Because memory accesses operation after detection operation do not affect detection, (3) and (4) merge into $\left\langle x \right\rangle Rx$.
    \item[3.]Formula \ref{label7}, When $O1$ exists, the positional relationship is $\left\langle x \right\rangle O1;Rx$. Memory accesses operation after sensitization may change the fault cell state, preventing detection. When $O1$ does not exist, the positional relationship is $\left\langle x \right\rangle ;Rx$. The test primitive formed by this relationship can detect State Fault.Analysis shows two test primitives, $\left\langle x \right\rangle Rx$ and $\left\langle x \right\rangle;Rx$, detect State Fault. Merging these yields $\left\langle x \right\rangle\#Rx$. Other test primitives cannot detect state faults. This proves the completeness of the State Fault test primitive.
\end{itemize}

The Read Destructive Fault occurs when a read operation inverts the fault cell state, reading the incorrect state. The read operation for sensitization detects this directly. The test primitive is $\left\langle Rx \right\rangle$, where the $D$ is $\phi$. Ultimately, there is only one $Rx$ operation. Its position and surrounding memory accesses operation are irrelevant. The final test primitive is $\left\langle Rx \right\rangle$, proving its completeness. The analysis of Incorrect Read Fault test primitive's completeness is similar and omitted here.

\subsection{Proof of Completeness for Double-cell Fault Coupled Testing Primitives}
To prove the completeness of double-cell coupled fault test primitives, this paper demonstrates three different scenarios:
\begin{itemize}
    \item [a.]The aggressor cell and victim cell are both in the state during sensitization.
    \item[b.]The aggressor cell is in the operation during sensitization.
    \item[c.] The victim cell is in the operation during sensitization.
\end{itemize}

\subsubsection{The aggressor cell and victim cell are both in the state during sensitization}

State Coupling Faults have the aggressor cell and victim cell both in the state during sensitization. The fault primitive is $\left\langle x;y/\overline{y}/- \right\rangle$, and the test primitive is shown in Table \ref{tab:table2}. Special cases have been introduced previously. Assume there are other forms of test primitives that can detect State Coupling Faults:
\begin{itemize}
    \item [1.] According to formula \ref{label5},When $O2$ exists,the positional relationships $;RyO2\left\langle x;y \right\rangle O3$  and $;RyO2\left\langle x;y \right\rangle $ form test primitives that can detect State Coupling Faults when the aggressor cell occupies the current address cell if x$\neq$y.When $O2$ does not exist,the positional relationships $;Ry\left\langle x;y \right\rangle O3$ and $;Ry\left\langle x;y \right\rangle $ form test primitives that can detect State Coupling Faults when the aggressor cell is in the CAC if $x=y$.Because the memory access operation following the detection operation does not impact fault detection,the test primitives formed by the above positional relationships can be summarized as follows:$\left\langle x;y \right\rangle$ \^{}$Ry*$.
    \item[2.]According to formula \ref{label6},When $O1$ exists,the positional relationships $\left\langle x;y \right\rangle O1RyO2$ and $\left\langle x;y \right\rangle O1Ry$  introduce multiple memory access operations following fault sensitization. If these operations alter the state of the fault cell, detection will fail.The positional relationships $\left\langle x;y \right\rangle RyO2$  and $\left\langle x;y \right\rangle Ry$  form test primitives that can detect State Coupling Faults when the victim cell occupies the CAC.Through analysis and synthesis,the test primitive $\left\langle x;y \right\rangle Ry$ is obtained.
    \item[3.] According to formula \ref{label7},When $O1$ exists, the positional relationship is $\left\langle x;y \right\rangle O1;Ry$. The test primitive formed based on this positional relationship adds multiple memory access operations after fault sensitization. When $O1$ does not exist, the positional relationship is $\left\langle x;y \right\rangle ;Ry$. The test primitive it forms can detect State Coupling Faults when the victim cell occupies the CAC if $x=y$. When x$\neq$y, the sensitization condition cannot be guaranteed because the memory array will be written to the same state before a March element starts.Through the above analysis, the formed positional relationships $\left\langle x;y \right\rangle Ry$ and $\left\langle x;y \right\rangle$ \^{}$Ry*$  are merged to obtain the test primitive:
    
\begin{equation}
\label{label10}
\left\langle x;y\right\rangle \left\{
    \begin{array}{cc}
         Ry& (v\enspace in\enspace CAC) \\
         \{^{}Ry*&(a\enspace in\enspace CAC) 
    \end{array}
\right.
\end{equation}
\end{itemize}

It can be seen that except for the form of the test primitive we have given, there are no other forms of test primitives that can detect State Coupling Faults, proving its completeness.

\subsubsection{The aggressor cell is in the operation during sensitization}

Disturb Coupling Fault CFdsrx has the aggressor cell sensitization as the operation during sensitization. Its fault primitive is $\left\langle Rx;y/\overline{y}/- \right\rangle$, and the test primitive can be obtained in Table \ref{tab:table2}. Assume there are other forms of test primitives that can detect CFdsrx. The analysis process is as follows:
\begin{itemize}
    \item [1.]According to formula \ref{label5}, when $O2$ exists, the positional relationships are $;RyO2\left\langle Rx;y \right\rangle O3$ and $;RyO2\left\langle Rx;y \right\rangle$. The test primitives formed based on these can detect CFdsrx in the a$\Rightarrow$v direction when x$\neq$y. The positional relationships when $O2$ does not exist are $;Ry\left\langle Rx;y \right\rangle O3$ and $;Ry\left\langle Rx;y \right\rangle$. The test primitives formed based on these can detect CFdsrx in the a$\Rightarrow$v direction when $x=y$.Organizing the test primitives formed by the above positional relationships, it is found that adding memory access operations after $Ry$ can detect CFdsrx in the a$\Rightarrow$v direction when x$\neq$ y, whereas not adding memory access operations can detect CFdsrx in the a$\Rightarrow$v direction when $x=y$. The ``*" can be add and obtain the test primitive. 
    \item[2.]According to formula \ref{label6}, four positional relationships can be obtained, but within these positional relationships, the sensitive operation and detection operation are in the same March element. In the current address cell, both the sensitive operation and detection operation act on the aggressor cell. However, when performing fault detection, it should be ensured that the detection operation acts on the victim cell. Therefore, the test primitives formed by these four positional relationships cannot detect CFdsrx faults.  
    \item[3.] According to formula \ref{label7},When $O1$ exists, the positional relationship at this time is $\left\langle Rx;y \right\rangle O1;Ry$. The test primitive formed based on this can detect CFdsrx faults in the v$\Rightarrow$a direction when x$\neq$y. When $O1$ does not exist, the positional relationship at this time is $\left\langle Rx;y \right\rangle ;Ry$. The test primitive formed based on this can detect CFdsrx faults in the v$\Rightarrow$a direction when $x=y$. The two positional relationships are merged together as $\left\langle Rx;y \right\rangle *;Ry$(v$\Rightarrow$a).Through the above analysis, the test primitives that can detect CFdsrx faults are obtained:
\begin{equation}
\label{label10}
\left\langle Rx;y\right\rangle \left\{
    \begin{array}{cc}
         \{^{}Ry*& (a\Rightarrow v) \\
         *;Ry&(v\Rightarrow a) 
    \end{array}
\right.
\end{equation}
\end{itemize}

It can be concluded that there are no other forms of test primitives that can detect this fault, proving the completeness of its test primitives. The completeness proofs of the two Disturb Coupling Fault test primitives CFdsxwx and CFdsxw!x are the same as the above type and will not be further proved in this paper.

\subsubsection{The victim cell is in the operation during sensitization}
The victim cell of Transition Coupling Fault is the operation during sensitization. Its fault primitive is $\left\langle x;yW \overline{y}/y/- \right\rangle$, and the test primitive is $\left\langle x;yW \overline{y} \right\rangle \#R\overline{y}$. Assume there are other forms of test primitives that can detect Transition Coupling Fault. Discuss according to the following conditions:

\begin{itemize}
    \item [1.]According to formula \ref{label5}, four positional relationships can be obtained. Because it cannot be guaranteed that fault detection temporally lags fault sensitization, the test primitives formed by these four positional relationships cannot detect Transition Coupling Fault.
    \item[2.] According to formula \ref{label6}, when $O1$ exists, the positional relationships are $\left\langle x;yW\overline{y} \right\rangle O1R\overline{y}O2$ and $\left\langle x;yW\overline{y} \right\rangle O1R\overline{y}$. After sensitization, there are multiple memory access operations that may change the state of the fault cells and cannot detect Transition Coupling Fault. The test primitives formed by the positional relationships $\left\langle x;yW\overline{y} \right\rangle R\overline{y}O2$ and $\left\langle x;yW\overline{y} \right\rangle R\overline{y}$ when $O1$ does not exist can detect Transition Coupling Fault. Because the memory access operations after the detection operation do not affect fault detection, the above positional relationships $\left\langle x;yW\overline{y} \right\rangle R\overline{y}O2$ and $\left\langle x;yW\overline{y} \right\rangle R\overline{y} $can be merged into $\left\langle x;yW\overline{y} \right\rangle R\overline{y}$.
    \item[3.] According to formula \ref{label7},When $O1$ exists, the positional relationship at this time is $\left\langle x;yW\overline{y} \right\rangle O1;R\overline{y}$. After sensitization, there are multiple memory access operations that may change the state of the fault cells and cannot detect Transition Coupling Fault. When $O1$ does not exist, the positional relationship at this time is $\left\langle x;yW\overline{y} \right\rangle ;R\overline{y}$. The test primitive formed based on this can detect Transition Coupling Fault.
\end{itemize}

Through the above analysis, the test primitives formed by the two positional relationships $\left\langle x;yW\overline{y} \right\rangle R\overline{y}$ and $\left\langle x;yW\overline{y} \right\rangle ;R\overline{y}$ respectively can detect Transition Coupling Fault. Merge them into the test primitive $\left\langle x;yW\overline{y} \right\rangle \#R\overline{y}$. It can be seen that except for the test primitive given in Table \ref{tab:table2}, there are no other forms of test primitives that can detect Transition Coupling Fault, proving its completeness. The completeness proofs of  Write Destructive Coupling Fault, Read Destructive Coupling Fault, Deceptive Read Destructive Coupling Fault and Incorrect Reading Coupling Fault test primitives are similar to that of  Transition Coupling Fault and will not be repeated here.

\section{The Application of Test Primitives }
The purpose of test primitives is to serve as a bridge between fault primitives and March algorithms to simplify the study of March algorithms. Test primitives describe the minimum test sequence corresponding to the fault primitives.In order to decouple cell status and fault sensitivity/detection,  state tuples are generated, based on the test primitives, to represent the state requirements for the minimum detection sequence to detect faults. At the same time, we also generate state tuples for March algorithms during their progress to describe the state of the entire memory array when reaching a certain memory access operation. On this basis, the test primitives and March algorithms are matched to complete the analysis of March algorithms. 
\subsection{Definition and Generation of State Tuples}
The state tuple of a test primitive represents the state conditions required for fault detection and the state tuple of a March algorithm represents the state of the memory array when the March algorithm proceeds to a given memory access operation. The expression of a state tuple is $\left\langle LAS/CAS/HAS\right\rangle$, where $LAS$ refers to the status of the low address cell, $CAS$ refers to the current address cell status, and $HAS$ refers to the high address cell status. $LAS, CAS,  HAS$  $\in$ \{0, 1, x\}, where 0 indicates that the memory cell state is 0, 1 indicates that the memory cell state is 1, and $x$ indicates that the memory cell state does not matter.

\subsubsection{Generation of  State Tuples in Test Primitive}

The generation of test primitive state tuples primarily depends on sensitizing the aggressor cell and victim cell. The steps to derive test primitives for coupled faults can be summarized as follows:  
\begin{itemize}
    \item [a.] Obtain the  sensitization operations  set from the test primitive;
    \item [b.]For sensitive operation sets that contain operations, the required status for those operations is placed in the current address cell's status . For sensitive operation sets without operations, any status is placed in the current address cell. Remove any sensitive conditions that have already been used in the current sensitive operation set. 
    \item[c.]According to the relative address positions of the aggressor cell and the victim cell, the remaining sensitization conditions in the sensitization operation set are used to determine the state of the high address cell or the low address cell, and the other element is set to $x$.   
    \item[d.]Generate the test primitive with state tuple by placing the state tuple before the sensitization operations and including the D detection operation.If only states are required, include only the state tuple and D detection operation in the final test primitive. 
\end{itemize}
For the state tuple of a single cell fault test primitive, simply change the current address cell status in the state tuple to the state required for fault sensitization. Set the high and low address cell status to $x$. The state required for the sensitization operation can be discussed in the following cases:
\begin{itemize}
    \item [1.]When the sensitization operation is a read operation, the state required for the sensitization operation is the expected state to be read out by this read operation, such as state 0 in r0.
    \item[2.] When the sensitization operation is a write operation, the state required for the sensitization operation is the state before the write operation is performed, such as state 0 in 0w1.
    \item[3.] When the sensitization operation is a state, the state required for the sensitization operation is the current state, such as state 0 in state 0. 
\end{itemize}

For the Transition Coupling Fault, when the aggressor cell is 0 and the victim cell is $1W0$ during fault sensitization, the test primitive is $\left\langle 0;1W0\right\rangle \#R0$. Its fault sensitization set can be obtained as $\left\langle 0;1W0 \right\rangle$. During fault sensitization, operations are required. Therefore, set $CAS$  in the state tuple to 1, representing the state required to sensitize the victim cell.Assuming the aggressor cell address is higher than the victim cell address, set $HAS$  in the state tuple to 1, representing the state required to sensitize the aggressor cell. Set $LAS$  to x . Therefore, the state tuple is $\left\langle x,1,1 \right\rangle$. The obtained test primitive with state tuple is $\left\langle x,1,1\right\rangle W0\#R0$.

\subsubsection{Generation of State Tuples in March Algorithms }

The generation of the entire state tuple of the March algorithm is done with March elements as cells. Generating the state tuple before the first memory access operation in each March element is relatively complex. From the second memory access operation onwards, this state tuple equals the state tuple following the previous operation. For conciseness, only the state tuple preceding each memory access operation is written. The state tuple following an operation depends on the operation type. A read operation does not alter the memory cell state, so the state tuple following equals that preceding. A write operation only modifies the current address, so the high and low address states in the tuple following remain unchanged. Only the current address state is rewritten to that following the write. For example, after w0 the current address state is 0.Here, two March element concepts are introduced: initial state and end state. In the March algorithm, from the second element onward the initial state equals the end state of the previous element. The initial state of the first element is either preset or $x$. The end state equals that following the final memory access operation. For example, there is a March element $\Uparrow(r0,r0)$,and the final operation is r0, the end state is 0.

Deriving the state tuple for the first memory access operation in a March element requires defining the initial and end states of that element. The following conclusions apply:  
\begin{itemize}
    \item [a.]For $\Uparrow$ address order, the state tuple is $\left\langle end \enspace state, initial \enspace state, initial \enspace state\right\rangle$.
    \item[b.] For $\Downarrow$ address order, the state tuple is $\left\langle initial \enspace state, initial \enspace state, end  \enspace state\right\rangle$.
    \item[c.] For $\Updownarrow$ address order, either tuple in a or b can apply.
\end{itemize}

\begin{figure}[!t]
\centering
\includegraphics[width=4in]{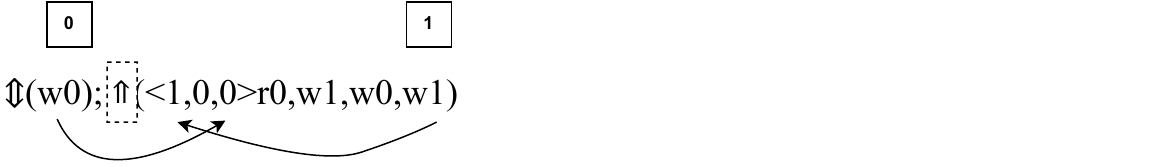}
\caption{State Tuple Generation Schematic}
\label{fig2}
\end{figure}

The expression for March A is: $\Updownarrow$(w0); $\Uparrow$(r0,w1,w0,w1); $\Uparrow$(r1,w0,w1); $\Downarrow$(r1,w0,w1,w0); $\Downarrow$(r0, w1,w0) . For the first element, the address order is $\Updownarrow$, the initial state is $x$, and the end state is 0. The state tuple is $\left\langle0,x,x\right\rangle$.
For the second element, the address order is $\Uparrow$. The initial state is the end state (0) of the first element. The end state is the state after the final operation w1, so is 1. The initial state tuple is $\left\langle1,0,0\right\rangle$.The state tuple generation schematic for the first memory access operation appears in Figure \ref{fig2}. Boxes above the March expression denote  end states of the March element. Lines below the expression, arrowed, indicate state tuple state sources.  Subsequent tuples is following state tuple after the previous operation. 
Combining the operations and tuples for the second element:
$\left\langle1,0,0\right\rangle\Uparrow r0\left\langle1,0,0\right\rangle\Uparrow w1\left\langle1,1,0\right\rangle\Uparrow w0\left\langle1,0,0\right\rangle\Uparrow w1$.Repeating this for each element in March A yields:
$\left\langle0,x,x\right\rangle\Updownarrow w0;\left\langle1,0,0\right\rangle\Uparrow r0\left\langle1,0,0\right\rangle\Uparrow w1\left\langle1,1,0\right\rangle\Uparrow w0\left\langle1,0,0\right\rangle 
\Uparrow w1; \left\langle1,1,1\right\rangle\Uparrow r1\left\langle1,1,1\right\rangle \Uparrow w0\left\langle1,0,1\right\rangle \Uparrow w1;\left\langle1,1,0\right\rangle\Downarrow r1\left\langle1,1,0\right\rangle      \Downarrow w0\left\langle1,0,0\right\rangle\Downarrow w1\left\langle1,1,0\right\rangle\Downarrow w0;\left\langle0,0,0\right\rangle\Downarrow r0\left\langle0,0,0\right\rangle\Downarrow w1\left\langle0,1,0\right\rangle
\Downarrow w0$.

\subsection{The Analysis Process of the March Algorithm}

\begin{figure}[!t]
\centering
\includegraphics[width=2.8in]{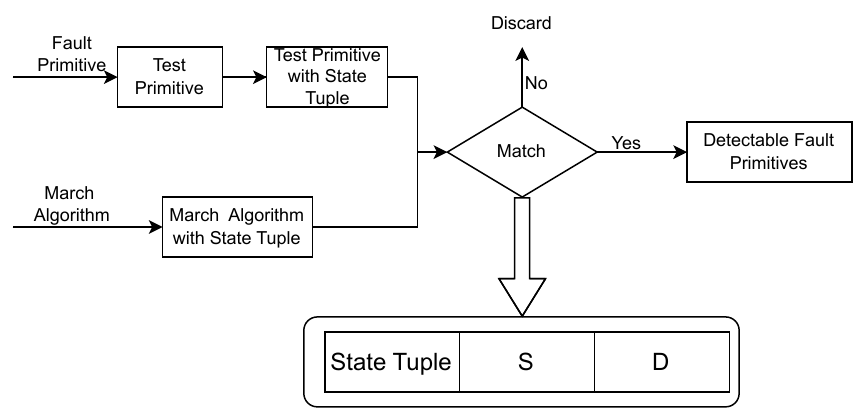}
\caption{Analysis Process Block Diagram}
\label{fig1}
\end{figure}

The March analysis process involves matching a test primitive with state tuples to a March expression with state tuples. If a match is found, then all faults detectable by the test primitive can also be detected by March. Matching a test primitive and March requires ensuring their state tuples, sensitized operation sets, and detection operation feature descriptor are compatible. The block diagram of the analysis process is shown in Figure \ref{fig1}.

For test primitive state tuples to match, non-x state values must remain unchanged in one March algorithm state tuple, while x-state values can be any state. To match the sensitized operation set, required operations and order must be the same. For the detection operation feature descriptor to match, the detection operation  and special symbols must be matched. As shown in Table \ref{tab:table1}, these symbols ``\^{}", ``\#", ``;"  denote the relative positional relationship between detection and sensitization operations. Matching symbols requires matching these relationships. Adding a memory access operation at ``*" that does not change the sensitized state is a match. 
Matching usually proceeds by first matching sensitized and detection operations, then state tuples, and finally special symbols. 

\begin{figure}[!t]
\centering
\includegraphics[width=3in]{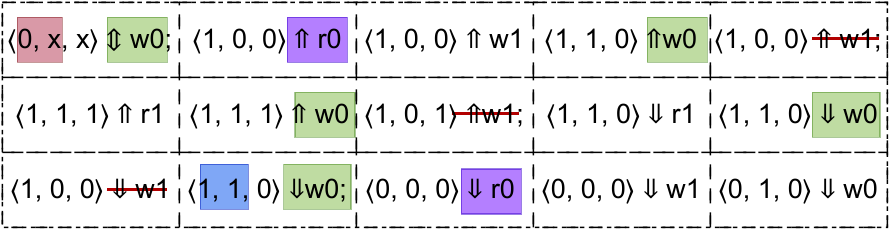}
\caption{State Primitive-March Algorithm Matching Schematic }
\label{fig3}
\end{figure}

For example, one Transition Coupling Fault primitive is $\left\langle 1;1W0/1/-\right\rangle$. Its test primitive is $\left\langle 0;1W0\right\rangle\#R0$. Assuming the aggressor cell address is lower than the victim cell address, the test primitive with state tuples is $\left\langle 1,1,x\right\rangle W0\#R0$.
For descriptive convenience, March A memory access operations 1 through 15 are numbered.As Figure \ref{fig3} shows,there are 5 memory access operations in each row. In it, green boxes denote matched sensitive operations; purple, matched detection operations; red lines, redundant memory access operations; red boxes, unmatched states; and blue boxes, matched states. Evidently, memory access operation 1 and the sensitive operation match but state tuples differ. Memory access operations 4,7,and 10 follow sensitive operations with redundant memory access operations, precluding matches. Memory access operation 12, followed by detection operation r0, has a state tuple that matches the above Transition Coupling Fault state tuple. The test primitive uses ``\#" to indicate the sensitization and detection operations can be in different March elements. In March A,operations 12 and 13 are in two elements,meeting the test primitive requirements. This test primitive matches March A. 
In the example,the sensitization operation is ``W0" and detection operation is ``R0". ``\#" indicates these operations can be separated in the March expression. Searching March A, operations 12 (``w0") and 13 (``r0") meet the requirements,with 12 as the sensitization operation and 13 as the detection operation. The state tuples  also match the test primitive state tuple $\left\langle1,1,x\right\rangle$. Therefore, this test primitive matches March A.

\subsubsection{Analysis of Fault Coverage for March A and March B}
The March A expression and state tuple expression were provided above. According to Table \ref{tab:table2},all non-connectivity static fault primitives and test primitives were obtained. Applying the method for obtaining state tuples yielded 86 test primitives with state tuples. These were matched with March A using state tuples and the matching method described previously. Fault primitives detectable by March A are summarized in Table \ref{tab:table3}, where each row denotes one fault model. Entries indicate March A fault primitives that can detect that model.  
For coupling faults,``a$>$v" signifies the aggressor cell address exceeds the victim cell address,``a$<$v" signifies the aggressor cell address is below the victim  cell address, and ``$\updownarrow$" signifies fault primitives in both directions are detectable. March A detects 44 of the 86 fault primitives.

\begin{table}[!t]
\caption{ Fault Primitives Detected by March A \label{tab:table3}}
\centering
\begin{tabular}{|c|c|}
\hline
Fault Model  &Detectable Fault Primitives  \\
\hline
SF	& \makecell{$\left\langle1/0/-\right\rangle$ ,$\left\langle0/1/-\right\rangle$}\\
\hline
 SAF	& \makecell[c]{$\left\langle {\forall}/0/-\right\rangle$ ,$\left\langle {\forall}/1/-\right\rangle$}\\
\hline
RDF & \makecell{$\left\langle R1/0/0\right\rangle$ ,$\left\langle R0/1/1\right\rangle$}\\
\hline
IRF & \makecell{$\left\langle R1/1/0\right\rangle$ ,$\left\langle R0/0/1\right\rangle$ }\\
\hline
TF & \makecell{$\left\langle 1W0/1/-\right\rangle$,$\left\langle 0W1/0/-\right\rangle$ }\\
\hline
CFst & \makecell{$\left\langle 0;0/1/-\right\rangle \updownarrow$ , $\left\langle 1;1/0/-\right\rangle\updownarrow$ , \\$\left\langle 0;1/0/-\right\rangle a> v$, $\left\langle 1;0/1/-\right\rangle a< v$  }\\
\hline
CFrd & \makecell{$\left\langle 0;R0/1/1\right\rangle\updownarrow$ , $\left\langle 1;R1/0/0\right\rangle\updownarrow$ , \\$\left\langle 0;R1/0/0\right\rangle a>v$ ,$\left\langle 1;R0/1/1\right\rangle a<v$ }\\
\hline
CFir & \makecell{$\left\langle 0;R0/1/1\right\rangle\updownarrow$ ,$\left\langle 1;R1/1/0\right\rangle\updownarrow$ ,\\ $\left\langle 0;R1/1/0\right\rangle a>v$,$\left\langle 1;R0/0/1\right\rangle a<v$}\\
\hline
CFdsrx & \makecell{$\left\langle R0;1/0/-\right\rangle a>v$, $\left\langle R1;0/1/-\right\rangle a < v$,\\$\left\langle R0;0/1/-\right\rangle\updownarrow$,$\left\langle R1;1/0/-\right\rangle\updownarrow$}\\
\hline
CFdsxw!x & \makecell{$\left\langle 0W1;0/1/-\right\rangle a<v$, $\left\langle 0W1;1/0/-\right\rangle a>v$, \\$\left\langle 1W0;0/1/-\right\rangle a<v$, $\left\langle 1W0;1/0/-\right\rangle\updownarrow$  }\\
\hline
CFtr & \makecell{$\left\langle 0;0W1/0/-\right\rangle a>v$,$\left\langle 0;1W0/1/-\right\rangle a>v$,\\$\left\langle 1;0W1/1/-\right\rangle\updownarrow$ ,$\left\langle 1;1W0/1/-\right\rangle a<v$}\\
\hline
\end{tabular}
\end{table}

Compared to March A, March B incorporates two additional read operations. Its expression is:
$\Updownarrow$(w0);$\Uparrow$(r0,w1,r1,w0,r0,w1);$\Uparrow$(r1,w0,w1);$\Downarrow$(r1,w0,w1,w0);\\$\Downarrow$(r0,w1,w0) .
The March B expression with state tuples, given in Table \ref{tab:table4}, was obtained  whereby each row denotes one March element. 

\begin{table}[!t]
\caption{ March B Expression with State Tuples \label{tab:table4}}
\centering
\begin{tabular}{|c|c|}
\hline
Number  &Expression  \\
\hline
1	& \makecell{$\left\langle 0,x,x\right\rangle \Updownarrow w0$}\\
\hline
 2& \makecell[c]{$\left\langle 1,0,0\right\rangle \Uparrow r0\left\langle 1,0,0\right\rangle  \Uparrow w1 \left\langle 1,1,0\right\rangle \Uparrow r0$ \\$\left\langle 1,1,0\right\rangle \Uparrow w0\left\langle 1,0,0\right\rangle  \Uparrow r0\left\langle 1,0,0\right\rangle  \Uparrow w1 $}\\
\hline
3& \makecell{$\left\langle 1,1,1\right\rangle \Uparrow r1 \left\langle 1,1,1\right\rangle \Uparrow w0 \left\langle 1,0,1\right\rangle \Uparrow w1$}\\
\hline
4 & \makecell{$\left\langle 1,1,0\right\rangle \Downarrow r1\left\langle 1,1,0\right\rangle  \Downarrow w0\left\langle 1,0,0\right\rangle  \Downarrow w1$ \\$\left\langle 1,1,0\right\rangle  \Downarrow w0$ }\\
\hline
5 & \makecell{$\left\langle 0,0,0\right\rangle \Downarrow r0\left\langle 0,0,0\right\rangle \Downarrow w1\left\langle 0,1,0\right\rangle \Downarrow w0$ }\\
\hline
\end{tabular}
\end{table}

Upon generating state-tuple test primitives from Table \ref{tab:table2}, these were matched with the March B expression with state tuples. The resulting fault primitives detectable by March B were obtained and statistically summarized in Table \ref{tab:table5}, whereby March B detects 47 of the 86 fault primitives. Notably, March B detects 3 additional  CFdsxw!x fault primitives compared to March A: $\left\langle 0W1;0/1/-\right\rangle a>v$,$\left\langle 1W0;0/1/- \right\rangle a>v$, and $\left\langle 0W1;1/0/- \right\rangle a<v$. Despite literature \cite{9} concluding equivalent fault coverage between March A and March B, this analysis finds March B exhibits slightly higher fault coverage for non-connectivity static  faults.

\begin{table}[!t]
\caption{ Fault Primitives Detected by March B \label{tab:table5}}
\centering
\begin{tabular}{|c|c|}
\hline
Fault Model  &Detectable Fault Primitives  \\
\hline
SF	& \makecell{$\left\langle1/0/-\right\rangle$ ,$\left\langle0/1/-\right\rangle$}\\
\hline
 SAF	& \makecell[c]{$\left\langle {\forall}/0/-\right\rangle$ ,$\left\langle {\forall}/1/-\right\rangle$}\\
\hline
RDF & \makecell{$\left\langle R1/0/0\right\rangle$ ,$\left\langle R0/1/1\right\rangle$}\\
\hline
IRF & \makecell{$\left\langle R1/1/0\right\rangle$ ,$\left\langle R0/0/1\right\rangle$ }\\
\hline
TF & \makecell{$\left\langle 1W0/1/-\right\rangle$,$\left\langle 0W1/0/-\right\rangle$ }\\
\hline
CFst & \makecell{$\left\langle 0;0/1/-\right\rangle \updownarrow$ , $\left\langle 1;1/0/-\right\rangle\updownarrow$ , \\$\left\langle 0;1/0/-\right\rangle a> v$, $\left\langle 1;0/1/-\right\rangle a< v$  }\\
\hline
CFrd & \makecell{$\left\langle 0;R0/1/1\right\rangle\updownarrow$ , $\left\langle 1;R1/0/0\right\rangle\updownarrow$ , \\$\left\langle 0;R1/0/0\right\rangle a>v$ ,$\left\langle 1;R0/1/1\right\rangle a<v$ }\\
\hline
CFir & \makecell{$\left\langle 0;R0/1/1\right\rangle\updownarrow$ ,$\left\langle 1;R1/1/0\right\rangle\updownarrow$ ,\\ $\left\langle 0;R1/1/0\right\rangle a>v$,$\left\langle 1;R0/0/1\right\rangle a<v$}\\
\hline
CFdsrx & \makecell{$\left\langle R0;1/0/-\right\rangle a>v$, $\left\langle R1;0/1/-\right\rangle a < v$,\\$\left\langle R0;0/1/-\right\rangle\updownarrow$,$\left\langle R1;1/0/-\right\rangle\updownarrow$}\\
\hline
CFdsxw!x & \makecell{$\left\langle 0W1;0/1/-\right\rangle \updownarrow$, $\left\langle 0W1;1/0/-\right\rangle \updownarrow$, \\$\left\langle 1W0;0/1/-\right\rangle \updownarrow$, $\left\langle 1W0;1/0/-\right\rangle\updownarrow$  }\\
\hline
CFtr & \makecell{$\left\langle 0;0W1/0/-\right\rangle a>v$,$\left\langle 0;1W0/1/-\right\rangle a>v$,\\$\left\langle 1;0W1/1/-\right\rangle\updownarrow$ ,$\left\langle 1;1W0/1/-\right\rangle a<v$}\\
\hline
\end{tabular}
\end{table}

\subsubsection{Analysis of Fault Coverage for March SR}
The expression for March SR is: 
$\Updownarrow(w0); \Uparrow(r0,w1,r1,w0); \Uparrow(r0,r0); \Uparrow(w1); \Downarrow(r1,w0,r0,w1); \Downarrow(r1,r1)$. 
Its state-tuple representation is:
$ \left\langle0,x,x\right\rangle \Updownarrow (w0);\left\langle 0,0,0\right\rangle  \Uparrow r0 \left\langle 0,0,0\right\rangle 
\Uparrow w1\left\langle 0,1,0\right\rangle 
\Uparrow r1\left\langle 0,1,0\right\rangle  \Uparrow w0;
\left\langle 0,0,0\right\rangle  \Uparrow r0\left\langle 0,0,0\right\rangle 
\Uparrow r0;
\left\langle 1,0,0\right\rangle  \Uparrow w1;
\left\langle 1,1,1\right\rangle  \Downarrow r1\left\langle 1,1,1\right\rangle   
\Downarrow w0\left\langle 1,0,1\right\rangle 
\Downarrow r0\left\langle 1,0,1\right\rangle 
\Downarrow w1;
\left\langle 1,1,1\right\rangle  \Downarrow r1\left\langle 1,1,1\right\rangle \Downarrow r1$. 
Upon matching these with state-tuple test primitives, the detectable fault primitives were obtained and summarized in Table \ref{tab:table6}, whereby March SR detects 62 of the 86 fault primitives.

According to literature \cite{10}, March SR can detect Write Destructive Faults but cannot detect Deceptive Read Destructive Fault. Write Destructive Faults requires writing without changing state, i.e., 0w0 or 1w1. However, March SR lacks such operations; thus, Write Destructive Faults cannot be sensitized and remain undetectable. In contrast, Deceptive Read Destructive Fault  necessitates two consecutive reads post-sensitization, which March SR can detect via dedicated test sequences. 
The test primitive for Write Destructive Faults is $\left\langle xWx\right\rangle\#Rx$. When x=0, the test primitive becomes $\left\langle 0W0\right\rangle \#R0$ with state-tuple $\left\langle x,0,x\right\rangle W0\#R0$. When x=1, the test primitive is $\left\langle 1W1\right\rangle \#R1$ with state-tuple $\left\langle x,1,x\right\rangle W1\#R1$. Numbering March SR's memory access operations from 1 to 14 reveals operations 1, 5 and 10 meet sensitization for x=0, yet their preceding state-tuples do not match. Operations 3, 8 and 12 satisfy sensitization for x=1 but also lack state-tuple matches, preventing Write Destructive Fault detection. 
The fault primitive for Deceptive Read Destructive Fault is $\left\langle Rx \right\rangle \#Rx$. When x=0 or x=1, the corresponding state-tuple test primitives are $\left\langle x,0,x\right\rangle R0\#R0$ and $\left\langle x,1,x\right\rangle R1\#R1$, respectively. Specifically, operations 6 and 7 enable Deceptive Read Destructive Fault detection for x=0 by satisfying sensitization and detection requirements with guaranteed state-tuples and symbol matching. Operations 13 and 14 similarly match the test primitive for x=1 Deceptive Read Destructive Fault. 
In summary, March SR cannot detect Write Destructive Faults but detects Deceptive Read Destructive Fault.

\begin{table}[!t]
\caption{ Fault Primitives Detected by March SR \label{tab:table6}}
\centering
\begin{tabular}{|c|c|}
\hline
Fault Model  &Detectable Fault Primitives  \\
\hline
SF	& \makecell{$\left\langle1/0/-\right\rangle$ ,$\left\langle0/1/-\right\rangle$}\\
\hline
 SAF	& \makecell[c]{$\left\langle {\forall}/0/-\right\rangle$ ,$\left\langle {\forall}/1/-\right\rangle$}\\
\hline
RDF & \makecell{$\left\langle R1/0/0\right\rangle$ ,$\left\langle R0/1/1\right\rangle$}\\
\hline
IRF & \makecell{$\left\langle R1/1/0\right\rangle$ ,$\left\langle R0/0/1\right\rangle$ }\\
\hline
TF & \makecell{$\left\langle 1W0/1/-\right\rangle$,$\left\langle 0W1/0/-\right\rangle$ }\\
\hline
DRDF & \makecell{$\left\langle R0/1/0 \right\rangle$,$\left\langle R1/0/1 \right\rangle$ }\\
\hline
CFst & \makecell{$\left\langle 0;0/1/-\right\rangle \updownarrow$ , $\left\langle 1;1/0/-\right\rangle\updownarrow$ , \\$\left\langle 0;1/0/-\right\rangle \updownarrow$, $\left\langle 1;0/1/-\right\rangle \updownarrow$  }\\
\hline
CFrd & \makecell{$\left\langle 0;R0/1/1\right\rangle\updownarrow$ , $\left\langle 1;R1/0/0\right\rangle\updownarrow$ , \\$\left\langle 0;R1/0/0\right\rangle \updownarrow$ ,$\left\langle 1;R0/1/1\right\rangle \updownarrow$ }\\
\hline
CFir & \makecell{$\left\langle 0;R0/1/1\right\rangle\updownarrow$ ,$\left\langle 1;R1/1/0\right\rangle\updownarrow$ ,\\ $\left\langle 0;R1/1/0\right\rangle \updownarrow$,$\left\langle 1;R0/0/1\right\rangle \updownarrow$}\\
\hline
CFdsrx & \makecell{$\left\langle R0;1/0/-\right\rangle \updownarrow$, $\left\langle R1;0/1/-\right\rangle \updownarrow$,\\$\left\langle R1;1/0/-\right\rangle\updownarrow$}\\
\hline
CFdsxw!x & \makecell{$\left\langle 0W1;0/1/-\right\rangle \updownarrow, \left\langle 1W0;0/1/-\right\rangle \updownarrow$, \\$\left\langle 1W0;1/0/-\right\rangle\updownarrow$  }\\
\hline
CFtr & \makecell{$\left\langle 0;0W1/0/-\right\rangle \updownarrow$,$\left\langle 0;1W0/1/-\right\rangle \updownarrow$,\\$\left\langle 1;0W1/1/-\right\rangle\updownarrow$ ,$\left\langle 1;1W0/1/-\right\rangle \updownarrow$}\\
\hline
CFdrd & \makecell{$\left\langle 0;R0/1/0\right\rangle \updownarrow$,$\left\langle 1;R1/0/1\right\rangle \updownarrow$}\\
\hline
\end{tabular}
\end{table}

\section{Conclusion}
This paper proposes a novel test primitive . The test primitive functions as a bridge between the fault primitive and the test algorithm, decoupling the cell state, fault sensitization, and detection and forming a highly flexible analysis unit. It is anticipated to streamline the analysis process of the March algorithm. For the set of faults addressed in this paper, a test primitive library is provided, and the completeness of the test primitives is demonstrated through proof by contradiction, thus ensuring completeness. The uniqueness and conciseness of the test primitives can also be guaranteed.The test primitives involved in this paper are applied in the analysis process of the March algorithm by expanding the state tuples. The algorithms such as March A, March B, and March SR have been analyzed. In the future, the test primitives will be used in automatically generating the March algorithm to achieve the purpose of simplifying this process.

\bibliographystyle{IEEEtran}
\bibliography{reference}










\newpage

 




\vfill

\end{document}